# Charged Momentum: Electric Vehicle Surge in India's 2023 Landscape

**Rahul Wagh[1]**

[1]Student, Dept. Of Mechanical Engineering, Vishwakarma Institute of Technology, Pune, Maharashtra, India

---------------------------------------------------------------------***---------------------------------------------------------------------

**Abstract -** *Electric vehicles (EVs) have emerged as a transformative force in India's transportation sector, offering a sustainable solution to the country's growing energy and environmental challenges. Against the backdrop of rapid urbanization, rising pollution levels, and the need for energy security, EVs have gained traction as a viable alternative to traditional internal combustion engine vehicles. This paper provides a comprehensive analysis of the electric vehicle market in India, focusing particularly on the landscape of 2023. It emphasizes key aspects such as the 2023 scenario of EV adoption, the role of indigenous manufacturers, dominant players shaping the market, and the influence of government policies and initiatives, including the FAME I and II schemes. Furthermore, the paper delves into EV sales data for the fiscal year 2023, offering insights into market trends and consumer preferences. By elucidating the current state of EVs in India, this paper aims to contribute to a deeper understanding of the country's transition towards sustainable mobility and its implications for energy, environment, and economy.*

**INTRODUCTION**

As of the fiscal year 2023, India has witnessed a substantial surge in cumulative Electric Vehicle (EV) sales, surpassed the 1 million mark in 2023 for the first time. The Society of Manufacturing of Electric Vehicles reports a noteworthy growth rate of 37.5%, highlighting the increasing popularity of electric vehicles in the country. Despite an initial ambitious target of achieving 100% EV adoption by 2030, the government has pragmatically revised this goal to 30%, showcasing a responsive approach to ground realities [1]. India's burgeoning transportation sector grapples with a significant gap between domestic crude oil production and consumption, as approximately 70% of its oil needs are met through imports [2]. This dependence poses environmental challenges, with a typical passenger vehicle contributing 4.7 metric tons of carbon emissions annually due to fossil fuel combustion [3].

In response, the Indian government envisions a strategic transition to electric vehicle production. This shift aims to substantially reduce the oil bill by US$60 billion and target a 37% reduction in emissions, fostering a decrease in reliance on fuel imports. Positioned as a protective measure against vulnerability to crude prices and currency fluctuations [4]

This transition is supported by a comprehensive array of government policies, including incentives, subsidies, and regulatory frameworks. The exploration of the multifaceted landscape of EV adoption in India extends to shedding light on the challenges faced by the electric vehicle industry, spanning infrastructural gaps to consumer concerns. Furthermore, the analysis delves into scrutinizing manufacturing trends and dissecting market dynamics, offering a comprehensive understanding of the factors influencing industry growth.

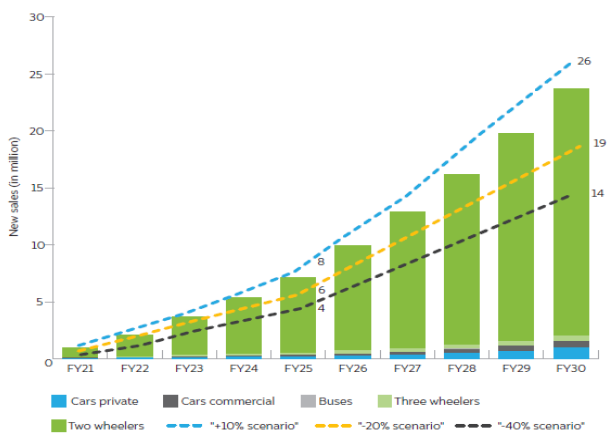

**Chart -1**:  Year to Year Rate of Increasing in EV'S

## 2023 SCENARIO OF ELECTRIC VEHICLES IN INDIA

The current landscape of electric vehicles (EVs) in India reflects a notable surge in growth and an evolving market. A pivotal study conducted by the Ola Mobility Institute in 2019 emphasized that achieving the central government's ambitious e-mobility targets for 2030 hinges on incentivizing EV adoption beyond the upfront purchase cost. The Total Cost of Ownership (TCO), encompassing direct and indirect expenses related to acquiring, operating, and maintaining EVs, remains a critical factor. Presently, the TCO for electric vehicles is comparatively higher due to various overheads, particularly given the nascent state of charging infrastructure in the country [6].

In the past year, the EV market has witnessed remarkable expansion, with a staggering 210% year-over-year increase. Notably, 999,949 EVs were sold in CY2022, a substantial rise from the 322,871 units in CY2021. The growth is predominantly attributed to the two- and three-wheeler segments, often considered the "close to the bottom peaches" of the EV business due to their affordability. This affordability factor positions them as the primary drivers of EV sales in the Indian market.

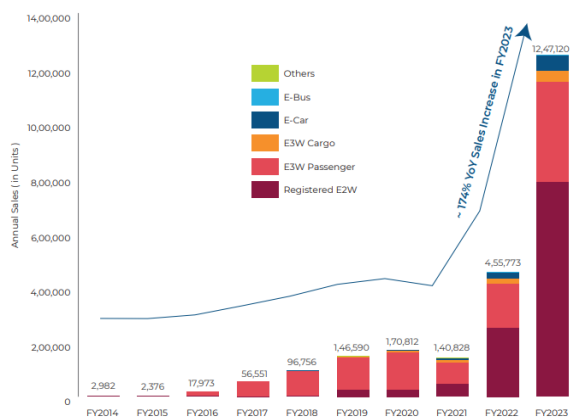

**Chart -2:** Growth of EV'S till 2023

Based on the findings of the 2023 Economic Survey, the domestic electric vehicle sector in India is forecasted to undergo a remarkable Compound Annual Growth Rate (CAGR) of 94.4% from 2022 to 2030. This growth trajectory is anticipated to culminate in 10 million EV sales annually by 2030, showcasing a substantial shift towards electric mobility. Furthermore, the sector is projected to generate an estimated 50 million direct and indirect jobs, indicating a significant economic impact on the nation's workforce.

## EV SALES FY2023

## E2W Market Share FY2023

The remarkable surge in electric vehicle (EV) sales has been primarily propelled by the electric two-wheelers (E2W) and three-wheelers (E3W) categories, jointly contributing to approximately 96% of the overall EV sales in the fiscal year. Notably, the E2W segment, known for its affordability, demonstrated robust growth, registering a substantial increase of 188% in total sales from 2,69,138 units in FY2022 to an impressive 7,74,614 units in FY2023. In this dynamic market, Ola Electric emerged as a frontrunner with an outstanding sales figure exceeding 1,50,000 units, followed closely by Hero Electric, which achieved remarkable sales of over 97,000 units in FY2023. Other key players such as Okinawa, TVS Motor, and Ampere Vehicles also made significant contributions, solidifying the overall success and dominance of the E2W segment in the electric vehicle landscape.

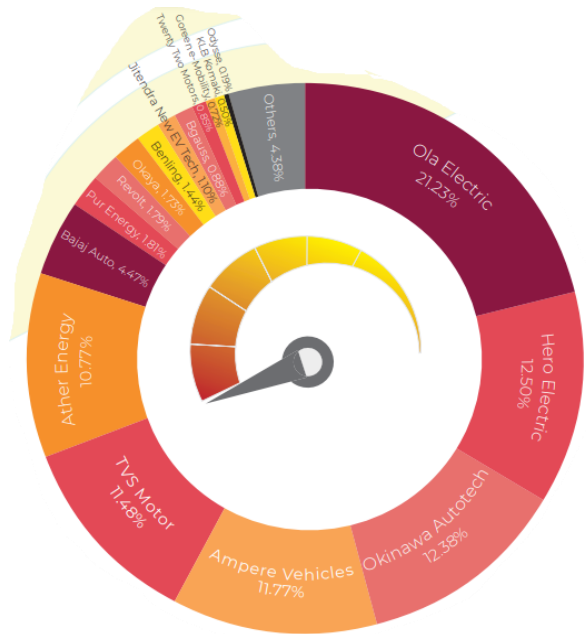

**Chart -3:** Key Players' Registered E2W Sales (FY2023)- (100% = 7,82,389 Units)

**E3W Market Share FY2023**

In the electric three-wheeler (E3W) category, the cumulative sales for FY2023 reached an impressive 4,07,099 units, marking a substantial growth compared to the 1,79,706 units recorded in FY2022. This notable year-on-year increase amounted to a robust 127% surge in sales. The landscape of E3Ws in the Indian market is characterized by diversity, and among the market leaders, Mahindra emerged at the forefront with sales reaching 35,675 units. Following closely is YC Electric, securing a substantial market share with 29,902 units sold. Saera Electric, Dilli Electric, and Champion Polyplast are noteworthy contenders in the E3W segment, contributing to the dynamic and evolving market for electric three-wheelers in India.

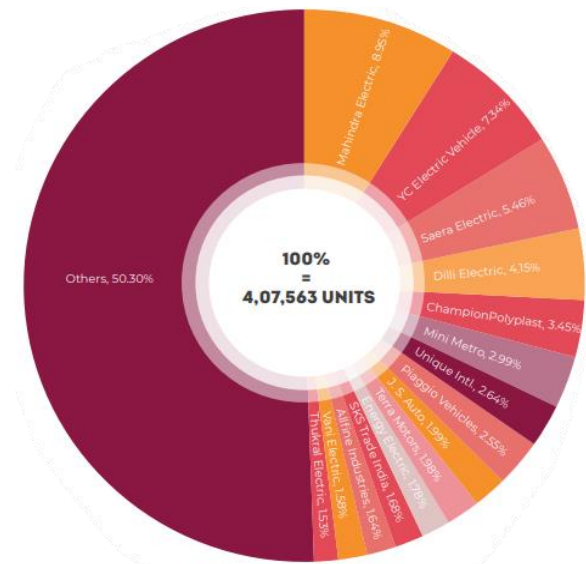

**Chart -4:** Player-wise E3W Sales' Share (FY2023)

**E4W Market Share FY2023**

In the electric car (E4W) segment, the overall sales for FY2023 reached an impressive 58,356 units, showcasing substantial growth from the 21,596 units recorded in FY2022. This remarkable surge translated to a significant 170% increase in sales year-on-year. Tata Motors has established itself as the dominant force in the market, commanding an impressive 80% market share. The company's sales figures for FY2023 stand at an impressive 47,006 units. Following Tata Motors, MG Motor and BYD India secured notable positions in the E4W segment with sales of 5,657 and 1,296 units, respectively, contributing to the diversification and growth of the electric car market in India.

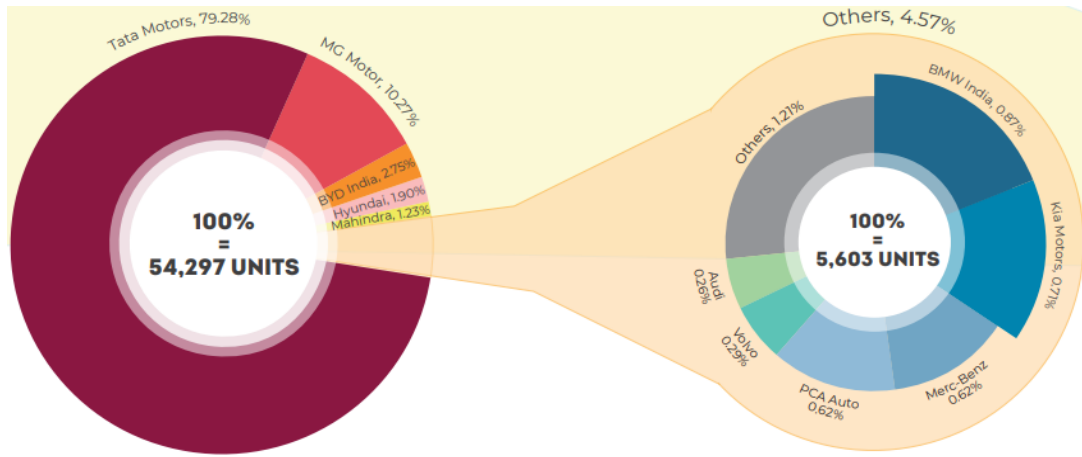

**Chart -5:** Player-wise E-Car Sales' Share (FY2023)

**E-BUS Market Share FY2023**

Within the electric bus (E-Bus) sector, the cumulative sales for FY2023 reached 1,913 units, a notable increase from the 1,186 units recorded in FY2022. This growth signifies a commendable 61% upswing in sales compared to the previous year. PMI Electro Mobility and Olectra collectively held a substantial market share, contributing to over 50% of the market, with sales figures of 604 and 444 units, respectively. Following closely are Switch Mobility/Ashok Leyland, JBM Auto, and Tata Motors, reinforcing the diverse landscape and growing adoption of electric buses in India.

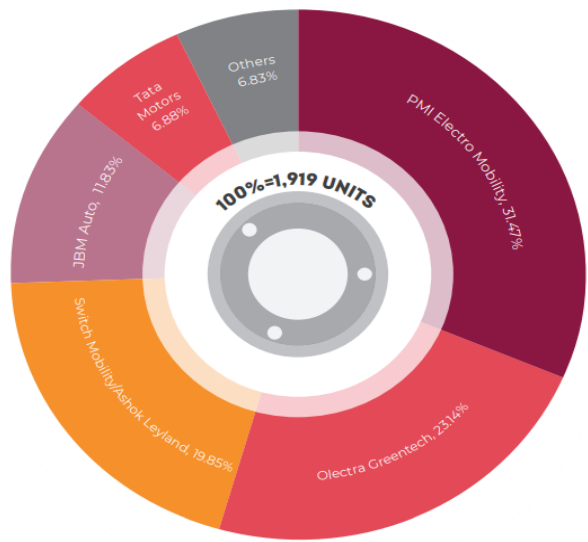

**Chart -6:** Player-wise E-Bus Sales' Share (FY2023)

**GOVERNMENT POLICIES AND INITIATIVES**

Over the past decade, the electric vehicle (EV) landscape in India has witnessed a dynamic evolution shaped by a series of initiatives and policy interventions. From 2011 to 2020, policymakers and regulators have played a pivotal role in steering the country towards sustainable mobility. The timeline during this period reflects a concerted effort to address challenges and capitalize on opportunities in the electric vehicle domain. This journey marks significant milestones, beginning with the inception of the National Electric Mobility Mission Plan (NEMMP) in 2013, followed by the initiation of the Faster Adoption and Manufacturing of Electric Vehicles (FAME) schemes, and strategic partnerships aimed at enhancing research and development in electric mobility. This timeline illustrates the proactive initiatives undertaken by authorities to incentivize the uptake of electric vehicles, mitigate emissions, and chart a path towards a more eco-friendly and sustainable transportation landscape in India.[15]

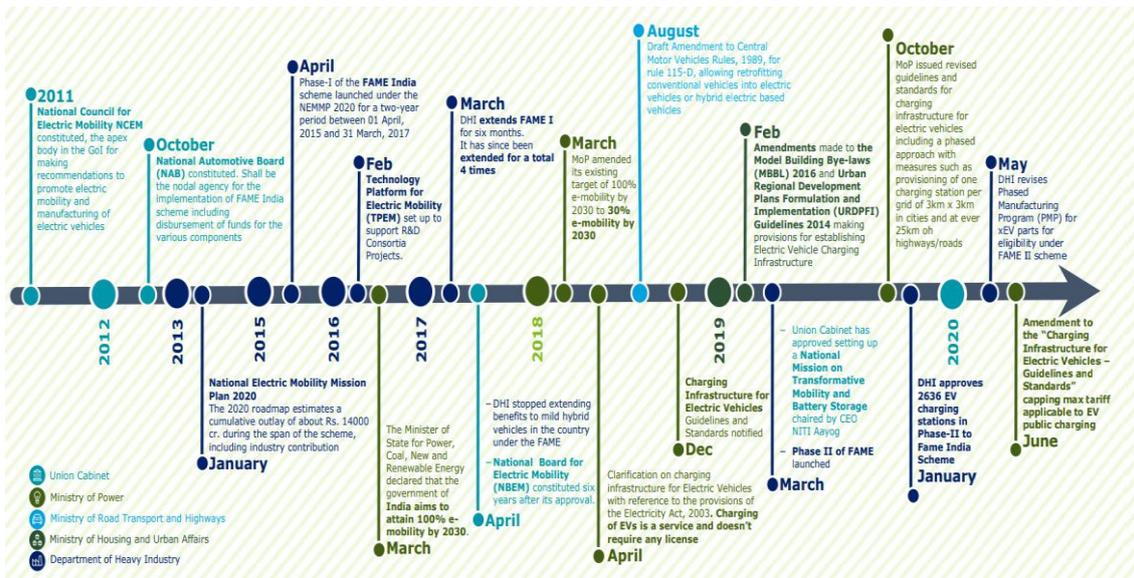

**Fig -1**: Timeline for various initiatives taken by policymakers and regulators 2011 to 2020

| Year | Initiatives | Schemes and Policies | Description |
|---|---|---|---|
| 2015-2017 | FAME-I | The Faster Adoption and Manufacturing of Electric Vehicles (FAME-I) was implemented with an initial budget allocation of INR 7.95 billion over a two-year period. | Introduced as part of NEMMP 2020. Focused on demand creation, technological platforms, pilot projects, and charging infrastructure. [13] |
| 2019-2022 | FAME-II India Programme | FAME II commits 10 billion rupees for developing charging infrastructure. | Allocates 10 billion rupees towards the development of charging infrastructure and authorizes the establishment of 241 recharge points dedicated to electric buses. [13] |
| 2020 | NEMMP 2020 | National Electric Mobility Mission Plan (NEMMP) 2020 aims for 5-7 million EVs by 2020. | Aims for 5-7 million EVs by 2020. Focuses on enhancing national energy security and domestic manufacturing capabilities. [8, 5] |

| Year | Scheme | Description | Details |
|---|---|---|---|
| 2021 | Production Linked Incentive (PLI) | Initiative started by the Government of India | The Scheme for the Automotive Sector, which was announced in September 2021 and has a budgetary allocation of INR 25,938 crore (approximately $3.1 billion), aims to boost domestic manufacturing of advanced automotive technology (AAT) products and stimulate investment across the automotive manufacturing value chain. |
| 2022-2030 | Green Mobility Roadmap | Proposed government initiative for the years 2022-2030 to encourage the transition to greener mobility solutions. | Aims to establish regulatory frameworks, incentives, and research projects to drive the adoption of green mobility. |
| 2023-2028 | State Government Initiatives | Karnataka and Maharashtra state-level initiatives promoting electric mobility. | Karnataka: Policy promoting research and development in electric mobility, mandating charging points in major cities. Maharashtra: Tax waivers for Electric Vehicles and establishment of India's first Electric Mass Mobility System. [8] |

**Table -1:** Year to year Schemes and Policies

**FAME I AND II SCHEME ROLE**

The FAME scheme stands as a cornerstone initiative by the Government of India, aimed at catalyzing the adoption of electric and hybrid vehicles across the nation. Designed to curtail the country's reliance on fossil fuels and foster sustainable transportation practices, the scheme has yielded significant success. Electric vehicle sales in India surged from 3,600 in 2013-14 to 156,000 in 2019-20, attributed largely to FAME's interventions. Spearheaded by the National Automotive Board (NAB), FAME has been instrumental in providing crucial incentives to both electric vehicle manufacturers and buyers. The scheme unfolds in two phases, with the second phase inaugurated in 2019, boasting a budget of Rs. 10,000 Crore over three years. Phase two is geared towards stimulating demand, earmarking support for 7,000 e-Buses, 500,000 e-3 Wheelers, 55,000 e-4 Wheeler Passenger Cars (including Strong Hybrid), and 1 million e-2 Wheelers.

The FAME scheme has additionally offered incentives to registered electric vehicle manufacturers and purchasers. As an integral component of the National Electric Mobility Mission Plan inaugurated in 2013 to advance electric mobility nationwide, the scheme has proven highly effective. Electric vehicle sales in India surged from 3,600 in 2013-14 to 156,000 in 2019-20, a testament to its success. Furthermore, the scheme has played a pivotal role in diminishing the country's carbon footprint while fostering sustainable transportation practices.

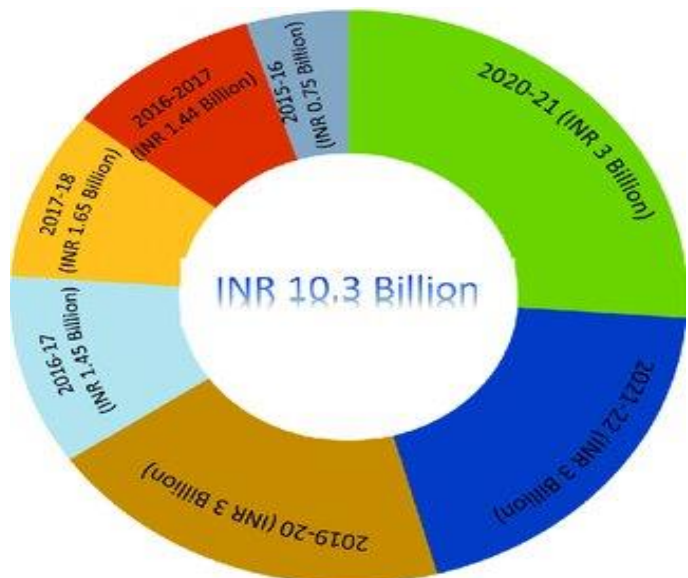

**Chart -7:** Fund allocation under the FAME scheme [13]

However, as per a report by The Economic Times, subsidies to certain electric two-wheeler manufacturers have been withheld due to alleged violations of local value addition requirements. Additionally, some manufacturers have faced accusations of underpricing their vehicles to qualify for subsidies, as highlighted in another article from Down To Earth. The government is reportedly considering legal action against EV manufacturers who have not returned wrongfully claimed incentives under FAME 2. These controversies have cast uncertainty over the future of FAME II, which may see resolution by the end of the current financial year. Nonetheless, despite these challenges, the FAME scheme has played a pivotal role in advancing the adoption of electric vehicles in India and mitigating the country's carbon footprint.

## MANUFACTURERS IN INDIA

As per the NITI Aayog government of India report, July 31, 2021, India witnessed a significant surge in the electric vehicle (EV) landscape with a remarkable presence of 380 electric vehicle manufacturers[17]. This proliferation reflects a robust and diverse ecosystem, showcasing the industry's dynamism and potential for growth. With the steady rise in the adoption of electric vehicles, this number is poised to expand further, affirming the nation's commitment to sustainable and eco-friendly mobility solutions. Notably, the Faster Adoption and Manufacturing of Electric Vehicles (FAME-II) initiative has played a pivotal role in shaping the landscape by approving various EV models from different manufacturers. In this context, the following chart provides insights into FAME-II approved electric vehicle models along with their respective manufacturers, offering a comprehensive overview of the evolving electric vehicle market in India.

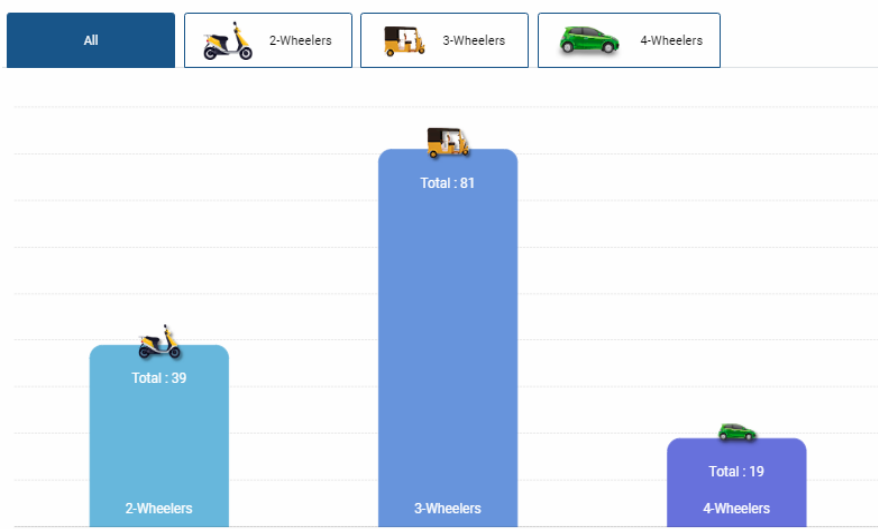

**Chart -8:** All types of vehicles

## 1. Electric 2-Wheelers[17]

As of July 31, 2021, the electric 2-wheeler segment in India has experienced substantial growth. Numerous manufacturers have entered this space, contributing to the diverse range of electric scooters and motorcycles available to consumers. FAME-II has played a crucial role in encouraging the adoption of electric 2-wheelers by approving various models from different manufacturers. With the rising popularity of electric 2-wheelers, this segment continues to be a key driver in the country's transition towards sustainable mobility.

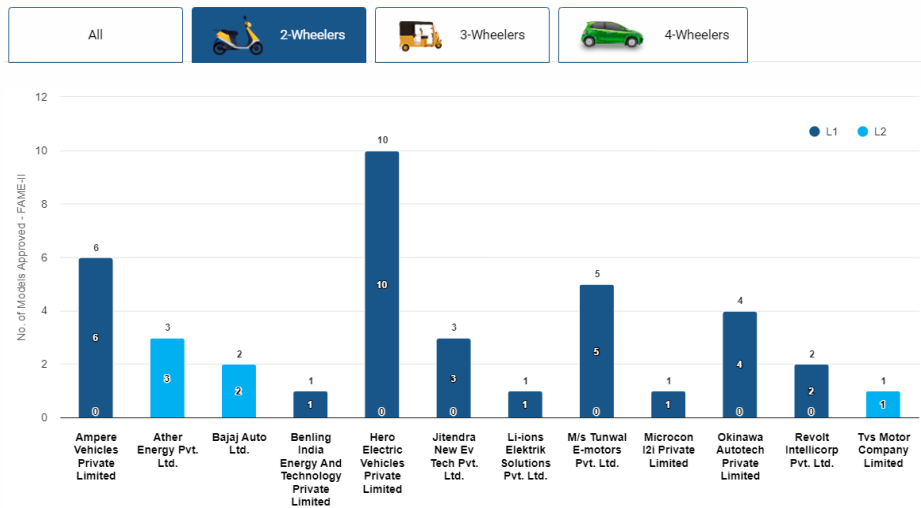

**Chart -9**: 2-Wheelers

## 2. Electric 3-Wheelers:[17]

The electric 3-wheeler segment has witnessed increased attention and adoption, particularly in the commercial and last-mile transportation sectors. FAME-II approval has been granted to several electric 3-wheeler models, showcasing the commitment of manufacturers to provide efficient and eco-friendly alternatives. The growth in electric 3-wheelers aligns with the goal of reducing emissions and enhancing the overall sustainability of urban transportation in India.

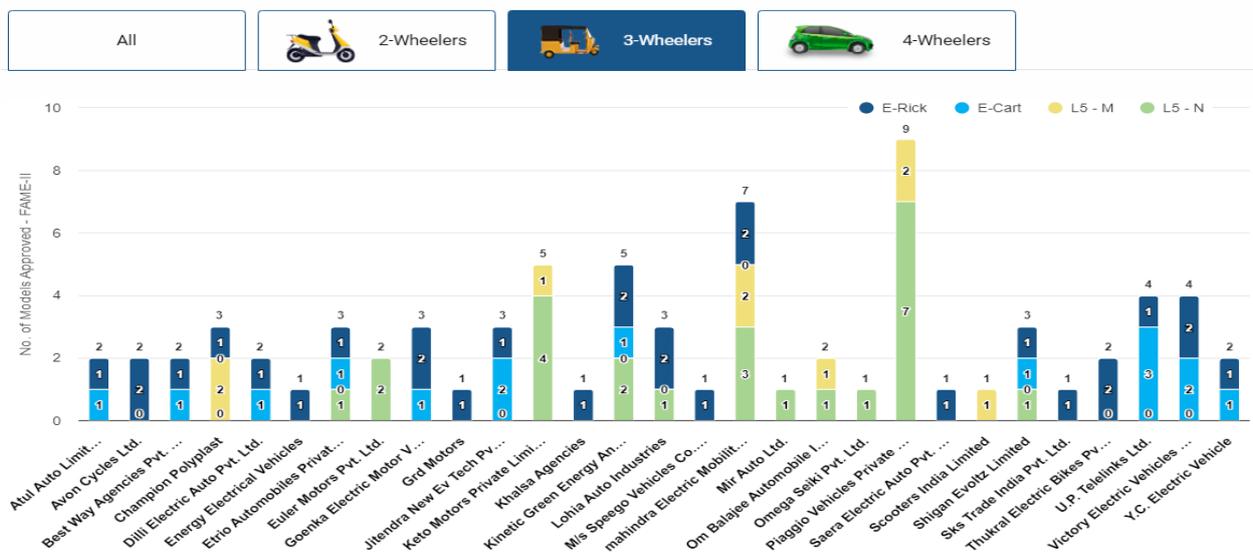

**Chart -10**: 3-Wheelers

## 3. Electric 4-Wheelers[17]

The electric 4-wheeler segment has seen notable developments with a variety of electric cars hitting the market. FAME-II's approval of electric 4-wheeler models underscores the industry's commitment to providing viable alternatives to traditional combustion-engine vehicles. Consumers now have an expanding choice of electric cars, reflecting a paradigm shift towards cleaner and greener mobility options.

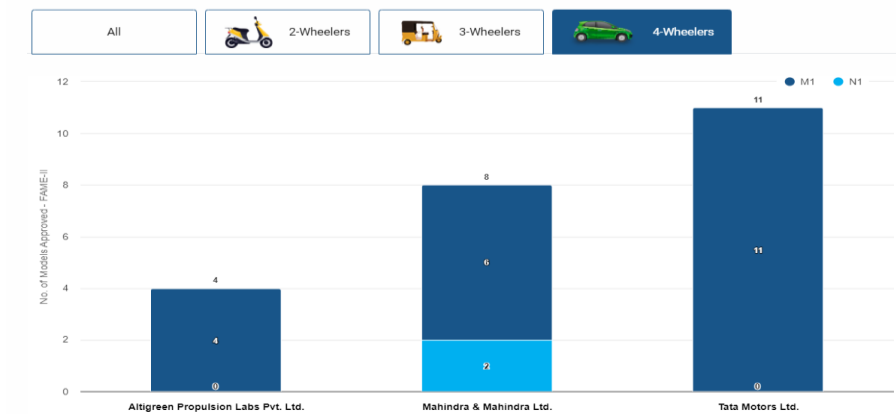

**Chart -11**: 4-Wheelers

## DOMINANT PLAYERS OF INDIA IN EV MARKET [14]

According to the most recent findings from Canalys research, Tata Motors emerged as the dominant player in the domestic EV market in 2023. The light vehicle market in India witnessed a 9.8% annual growth in the first half of 2023, reaching a total of two million units. Within this growth, EVs contributed to approximately 2.4%, experiencing an impressive annual growth rate of around 137%.

- Market Share of Indian Auto Brands in EVs (H1 2023):
    - Tata Motors: Leads with a 42% share.
    - Hyundai: Follows with 15%.
    - MG: Holds 14%.
    - Mahindra: Has 10%.
    - Kia: Represents 7%.
    - Toyota and Others: Make up the remaining shares.
- Battery Electric Vehicle (BEV) Market Share in India (H1 2023):
    - Total market: Around 48,000 units.
    - Tata Motors: Sold approximately 34,000 units.

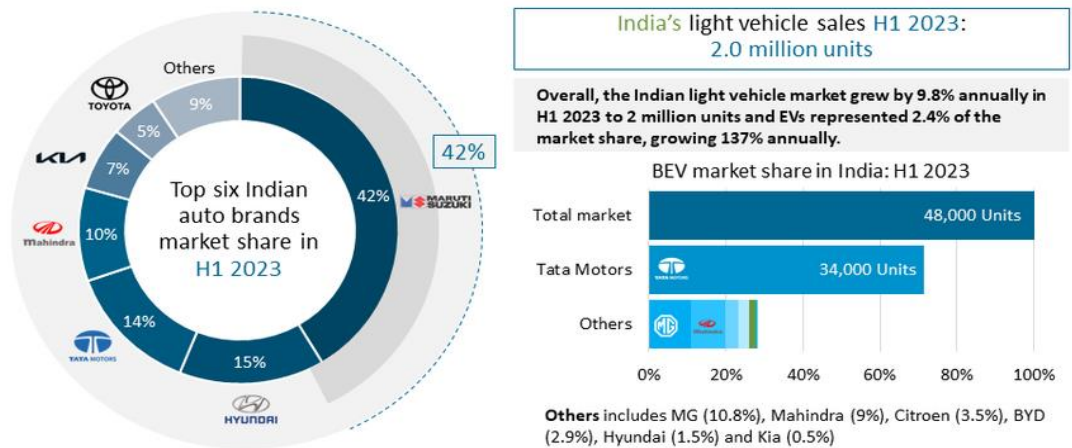

**Chart -12**: Tata leads EVs, with MG and Mahindra catching up

India's automotive market has surged in recent years, driven by a concerted effort to transition to electric vehicles (EVs) as part of the government's ambitious targets. With OEMs aligning their strategies towards EVs and the introduction of numerous models, the market has witnessed significant growth, particularly in the first half of 2023. Tata Motors has dominated the scene with its affordable yet feature-rich offerings, capturing a substantial market share. Other players like MG, Mahindra, and Citroen have also made notable strides, contributing to the diversification of the EV market. Despite challenges and competition, Canalys anticipates continued growth, projecting the EV market to expand to over 300,000 units in 2025, representing a substantial share of the overall light vehicle market.

Furthermore, the landscape is evolving with new entrants like Maruti Suzuki and Ola joining the fray, promising to further accelerate the adoption of EVs. While global giants like Tesla are eyeing the Indian market, local manufacturing and supply chain development remain critical factors. Government incentives and policies, including the FAME India Phase II scheme, have provided impetus, but there's a pressing need for a robust ecosystem of EV component suppliers to sustain and propel this growth trajectory. As India emerges as a pivotal market for EVs, leveraging its talent pool, supportive policies, and active subsidies, the stage is set for a dynamic and competitive EV market landscape in the years to come.

## CONCLUSIONS

The evolution of electric mobility in India marks a transformative journey toward sustainable transportation solutions. As the nation grapples with urbanization and environmental challenges, the embrace of electric vehicles (EVs) emerges as a pivotal strategy to mitigate carbon emissions, reduce dependency on fossil fuels, and foster economic growth. The fiscal year 2023 witnessed a significant milestone, surpassing the 1 million mark in cumulative EV sales, fueled by robust growth rates and government interventions.

Key players in the EV market, particularly in segments like electric two-wheelers (E2W), three-wheelers (E3W), and four-wheelers (E4W), have demonstrated remarkable strides, buoyed by supportive policies and incentives such as the Faster Adoption and Manufacturing of Electric Vehicles (FAME) schemes. Tata Motors has emerged as a dominant force, spearheading the transition with innovative offerings and capturing a substantial market share.

Government initiatives, ranging from the National Electric Mobility Mission Plan (NEMMP) to the Production Linked Incentive (PLI) scheme, underscore a concerted effort to nurture the EV ecosystem, bolstering research and development, charging infrastructure, and domestic manufacturing capabilities. Despite challenges, including controversies surrounding subsidy disbursements, the trajectory of electric mobility remains optimistic, with projections indicating exponential growth in EV sales and job creation by 2030. Looking ahead, India stands at the cusp of a dynamic and competitive EV market landscape, with new entrants like Maruti Suzuki and Ola poised to drive further adoption. Collaboration between stakeholders, innovation in technology, and a conducive regulatory framework will be critical in realizing the vision of a greener, cleaner, and more sustainable future for transportation in India. As the world's third-largest emitter of greenhouse gasses, India's commitment to electric mobility not only aligns with global sustainability goals but also presents opportunities for economic prosperity and environmental stewardship on a national scale.